\documentclass[lettersize,journal]{IEEEtran}

\usepackage[T1]{fontenc}
\usepackage[dvipsnames,svgnames,x11names]{xcolor}
\usepackage[noblocks]{authblk}

\usepackage[utf8]{inputenc}
\usepackage[nolist]{acronym}
\usepackage{graphicx}
\usepackage[export]{adjustbox}
\usepackage{float}
\usepackage{tikz}
\usepackage{framed}
\usepackage{comment}
\usepackage{gensymb}
\usepackage{multirow}
\usepackage{hyperref} 
\usepackage{url}
\usepackage{footmisc}
\usepackage{siunitx}
\usepackage{threeparttable} 
\usepackage[table,dvipsnames]{xcolor}
\colorlet{lightred}{red!30!white}
\colorlet{lightorange}{orange!30!white}
\colorlet{lightgreen}{green!30!white}
\colorlet{lightblue}{blue!30!white}
\colorlet{lightviolet}{violet!30!white} 
\colorlet{lightcyan}{cyan!30!white}
\usepackage{booktabs} 
\usepackage{caption}
\usepackage{svg}
\usepackage{diagbox}
\usepackage[english]{babel}

\usepackage{cite}

\ifCLASSINFOpdf
\else
\fi
\usepackage{amsmath}
\usepackage[cmintegrals]{newtxmath}
\usepackage{bm}

\usepackage{datetime}
\usepackage{caption}

\ifCLASSOPTIONcompsoc
  \usepackage[caption=false,font=normalsize,labelfont=sf,textfont=sf,subrefformat=parens,labelformat=parens]{subfig}
\else
  \usepackage[caption=true,font=footnotesize,subrefformat=parens,labelformat=parens]{subfig}
\fi
\usepackage{atbegshi}
\AtBeginShipoutFirst{\copyrightnotice}

\newcommand\copyrighttext{%
  \footnotesize \textbf{This work has been submitted to an IEEE Journal for possible publication. Copyright may be transferred without notice, after which this version may no longer be accessible.}}

\usepackage{everypage}

\newcommand\copyrightnotice{%
\begin{tikzpicture}[remember picture,overlay]
\node[anchor=north,yshift=-4pt] at (current page.north) {\fbox{\parbox{\dimexpr\textwidth-\fboxsep-\fboxrule\relax}{\copyrighttext}}};
\end{tikzpicture}%
}

\begin{document}
\pagestyle{empty} 

\title{Threshold Optimization and Dynamic Adaptation of Distributed Optimal Power Flow in 5G Networks}


\author[*]{Biswajit Kumar Dash}
\author[*]{Garrett Thomas}
\author[*]{Adedoyin Inaolaji}
\author[*]{Filippo Malandra}
\affil[*]{Department of Electrical Engineering, State University of New York at Buffalo, NY, USA}
\renewcommand\Authands{, and }



\maketitle
\thispagestyle{empty} 
\begin{acronym}[WInnForum] 
\acro{Acronyms}{}
\acro{AP}{Access Point}
\acro{ADMM}{Alternating Direction Method of Multipliers}
\acro{APP}{Auxiliary Principle Problem}
\acro{ATC}{Analytic Target Cascading}
\acro{ALADIN}{Augmented Lagrangian Based Alternating Direct Inexact Newton}
\acro{BS}{Base Station}
\acro{CI}{Confidence Interval}
\acro{CHIL}{Controller-Hardware-in-the-Loop}
\acro{DOPF}{Distributed Optimal Power Flow}
\acro{DER}{Distributed Energy Resources}
\acro{GU}{Global Update} 
\acro{HIL}{Hardware-in-the-Loop}
\acro{KPI}{Key Performance Indicator}
\acro{LC}{Local Controller} 
\acro{LU}{Local Update}
\acro{NTP}{Network Time Protocol}
\acro{OPF}{Optimal Power Flow}
\acro{PV}{Photovoltaic}
\acro{RTT}{Round-Trip Time}
\acro{RSRP}{Reference Signal Received Power}
\acro{RPi}{Raspberry Pi}
\acro{RSRQ}{Reference Signal Received Quality}
\acro{SINR}{Signal-to-Interference-plus-Noise Ratio}
\acro{UDP}{User Datagram Protocol}
\acro{SG}{Smart Grid}
\acro{RMSE}{Root Mean Square Error}
\acro{Confidence interval}{CI}

\end{acronym}

\begin{abstract}
In this paper, we present an experimental evaluation study of the Alternating Direction Method of Multipliers (ADMM), which is a widely used technique in the distributed optimization of power distribution networks. The focus of this study is on how real 5G communication performance affects ADMM in a fully experimental platform that features commercial 5G connectivity and real-time control. The ADMM-based Distributed Optimal Power Flow (DOPF) problem is solved using the IEEE 123-bus unbalanced distribution feeder subdivided into five areas, each managed by a local controller implemented on a Raspberry Pi.
To mitigate the impact of the communication network variability, we propose a delay threshold-based mechanism that yields a 7.75\% reduction in convergence time compared to a no-threshold baseline.
We also devised a policy to dynamically update the threshold value based on communication and computation conditions, achieving a 26.42\% reduction in the convergence time compared with the static optimal threshold. These results demonstrate the potential of adaptive, communication-aware control strategies for real-world Smart Grid (SG) deployments.
\end{abstract}


\IEEEpeerreviewmaketitle


\section{Introduction}\label{section:introduction}

\IEEEPARstart{W}{ith} growing concerns over the environmental impacts of fossil-fuel generation and global commitments to reach net-zero emissions by 2050, renewable energy integration is expanding rapidly. The increasing penetration of \acp{DER} is transforming the operation of power distribution networks. The large-scale integration of renewable energy and widespread \acs{DER} deployment introduces new operational challenges for traditional centralized control architectures, which suffer from scalability limitations, high computational burdens on a single controller, single points of failure, and cybersecurity and privacy concerns regarding local data
. Given these challenges, distributed optimization serves as an alternative approach for enabling decentralized control of \acp{DER}~\cite{ochoa2017distributed, schneider2019distributed}.


The main rationale behind distributed optimization methods is to decompose large-scale optimization problems into smaller subproblems that \acp{LC} can solve locally using local data~\cite{baldick2002fast}. This approach enhances the resilience of power grids by distributing computational loads, reducing vulnerability to cyber-physical attacks, and improving privacy compared to centralized methods. Among distributed optimization techniques, the \ac{ADMM} has emerged as one of the most widely adopted methods for power system applications~\cite{zheng2015fully}. ADMM combines the decomposability of dual ascent with the robustness of augmented Lagrangian methods, enabling independent \acp{LU} followed by consensus exchanges with a central coordinator~\cite{boyd2011distributed}. This structure makes ADMM attractive for \ac{OPF}, which is used to optimize a chosen system objective, such as minimizing voltage drop across a system or total cost of generation, by setting the active and reactive power setpoints of \acp{DER}. Such a distributed approach offers strong convergence guarantees, scalability to large networks, and the ability to preserve local data privacy.

The performance of \ac{ADMM} is typically characterized by several key metrics: the number of iterations required for convergence, the maximum primal residual (i.e., the consensus constraint violation between local and global variables), the objective value relative to the centralized solution, and the computation and communication delays experienced by \acp{LC}. Because the \acs{ADMM} relies on iterative information exchange among \acp{LC} and the central coordinator, the performance of the underlying communication network directly influences convergence speed and overall effectiveness. Therefore, accurately modeling communication network behavior is essential for the realistic evaluation of \acs{ADMM} performance and \ac{SG} system design.

In standard \ac{ADMM}, the coordinator must receive updates from all \acp{LC} before proceeding~\cite{boyd2011distributed}, which can cause deadlock under delays or packet loss. To account for network delay and its impact on \ac{ADMM} performance, in our prior work~\cite{inaolaji2024analyzing}, we proposed a mechanism that allows the coordinator to proceed upon the expiration of a preset delay threshold, regardless of the number of updates received from the \acp{LC}. However, the evaluation of this threshold-based update policy in \cite{inaolaji2024analyzing} was based on analytical delay models that fail to capture the real-world variability of communication networks.

To address this gap, we developed ExTODS (Experimental Testbed for Optimization of Distributed Systems)~\cite{dash2025extods}, a fully experimental testbed that features bidirectional 5G communications between LCs and a control center. ExTODS is used to solve an \ac{ADMM}-based \ac{DOPF} problem, with each LC implemented on a \ac{RPi} that communicates over a commercial 5G network. It is worth remarking that ExTODS, differently from~\cite{inaolaji2024analyzing}, does not rely on statistical delay distribution, but instead allows to experimentally measure the network latency.

In this work, the \ac{ADMM} algorithm is experimentally evaluated by partitioning the IEEE 123-bus feeder into five areas using a spectral clustering method. We have investigated how 
the delay threshold 
affects \ac{ADMM} network performance, and we proposed a novel technique to determine the optimal delay threshold both experimentally and analytically. 

The main contributions of this work are:

\begin{itemize}
  
  \item We scale ExTODS to a five-area partition of the IEEE 123-bus feeder and conduct comprehensive experiments on a real-world cellular network to characterize the impact of communication delays on \ac{ADMM} convergence behavior, iteration count, and computation time.
  
    \item We experimentally derive an expression for the optimal static threshold that balances the \ac{ADMM} algorithm's computational efficiency and solution accuracy.
    
    \item We propose a novel dynamic threshold selection strategy that adapts in real time to changing network conditions using measured \acp{RTT} and \acs{LC} optimization times, and demonstrate that this approach consistently outperforms static thresholding under variable communication environments.

   %
  
\end{itemize}

The rest of this paper is organized as follows. Section~\ref{section:RelatedWork} presents related work. Section~\ref{section:SystemModel} describes the architecture of ExTODS, the power system model, and the \acs{ADMM} algorithm. Section~\ref{section:Setup} presents the experimental setup. Section~\ref{section:Static_Threshold} discusses the \ac{ADMM} performance under static thresholds and derives the optimal threshold expression. Section~\ref{section:Dynamic_Threshold} presents the dynamic threshold tuning framework and corresponding experimental results. Finally, Section~\ref{section:Conclusion} concludes the paper.

\section{Related Work}\label{section:RelatedWork}

Distributed optimization has been widely studied to address scalability, privacy, and resilience challenges in modern distribution grids. A comprehensive survey of distributed optimization algorithms for power systems is provided in~\cite{molzahn2017survey}. Several well-known techniques---including \ac{ADMM}~\cite{zheng2015fully}, \ac{ATC}~\cite{kargarian2014distributed}, \ac{ALADIN}~\cite{engelmann2017distributed}, and \ac{APP}~\cite{baldick1995generalized}--have been evaluated for solving \acs{DOPF} problems. Recent works have also explored algorithmic enhancements to improve \ac{ADMM}'s convergence behavior. For example, Hasanzadeh and Kargarian~\cite{hasanzadeh2025distributed} proposed three \ac{ADMM}-based techniques to accelerate convergence for \acs{DOPF} problems. 
However, these approaches primarily focus on algorithmic improvements and are evaluated under idealized or simulated communication conditions.

In practice, the performance of distributed optimization is tightly coupled with the characteristics of the underlying communication network. Prior studies have examined communication-aware aspects of distributed control: \cite{zhang2012convergence} and \cite{mohammadi2014role} analyzed the impact of communication topology on consensus and economic dispatch, assuming ideal communications with no delay, packet loss, or bandwidth constraints. To account for less-than-ideal communication,~\cite{tsianos2011distributed} studied the effects of fixed and random delays on distributed optimization and proposed delay models to better capture network behavior. Still, many practical communication impairments, such as packet loss, link failures, interference, or fluctuating \ac{SINR}, remain underexplored in distributed optimization frameworks.


Recent studies have sought to incorporate more realistic wireless communication characteristics into distributed control and optimization frameworks. In \cite{liu2017distributed}, a neighbor-based voltage control strategy was validated under analytically modeled random communication link failures, where link availability follows a stochastic process. The work in~\cite{tian2025frequency} developed a fault-tolerant \ac{ADMM} scheme that reuses the most recent available data during communication failures, where communication failures are modeled analytically using stochastic link-failure processes. A closely related fault-tolerant, threshold-based update mechanism was introduced in our earlier work \cite{inaolaji2024analyzing}, in which communication delays are modeled analytically using delay distributions from the literature. Limited-bandwidth and time-varying networks were modeled analytically in \cite{li2017distributed} using distributed subgradient methods over time-varying graphs and validated through numerical simulations. Guo et al. \cite{guo2017role} compared centralized and distributed implementations of \ac{ADMM} and Optimality Condition Decomposition (OCD) using the OPNET simulation suite. Additional approaches to mitigating communication imperfections in distributed voltage control include freezing mechanisms for analytically modeled stochastic link failures~\cite{li2020distributed} and weighted autoregressive updates to accelerate convergence under analytically modeled random communication delays~\cite{xu2018admm}. However, across these studies, communication impairments are either abstracted analytically (e.g., stochastic link failures or delay distributions) or emulated through simulation platforms (e.g., OPNET), and the distributed algorithms are validated primarily through numerical simulations.


Complementary to the studies discussed above, several works have explored experimental and co-simulation platforms that integrate power-system control with communication networks. In particular, some platforms combine physical power-system hardware components with simulated communication networks~\cite{hopkinson2006epochs, lin2012geco}. Other implementations employ real communication networks using protocols such as DNP3 and Modbus TCP~\cite{khan2021attack}, as well as Ethernet and JTAG~\cite{kumar2025real}. However, these platforms are used for testing protection schemes and cybersecurity, rather than distributed optimization. 
Shah et al.~\cite{shah2023experimental}, and Dash et al.~\cite{dash2023network} measured \acs{SG} communication delays over low-power cellular technologies. However, these technologies and the measured delays did not prove suitable for delay-sensitive applications such as distributed optimization in the distribution system. Moreover, those measurements were taken using testbeds that evaluated only communication network performance, without integrating a power-system optimization framework. In~\cite{gebbran2020practical}, an \acs{RPi}-based \acs{DOPF} prototype was built to coordinate \acs{DER}, but only part of the prosumer subproblems ran in parallel; the rest were solved sequentially on a PC over a Wi-Fi network. The \ac{CHIL} testbed in~\cite{kharchouf2024controller} focused on distributed consensus under cyberattacks over synchronized Wi-Fi links without evaluating distributed optimization.

Differently from the reviewed body of literature, which either relies on analytically abstracted or simulated communication models, or focuses on experimental platforms without integrated distributed optimization, this paper presents an experimental evaluation of \acs{ADMM} that explicitly considers power system dynamics while using a real communication network to transmit and receive data packets. In addition, this paper proposes a strategy to mitigate the impact of communication network performance degradation on \ac{ADMM} performance in a time-varying scenario.


\section{System Model}\label{section:SystemModel}
\begin{figure}[!tbp]
    \centering
    \centerline{\includegraphics[width=0.5\textwidth]{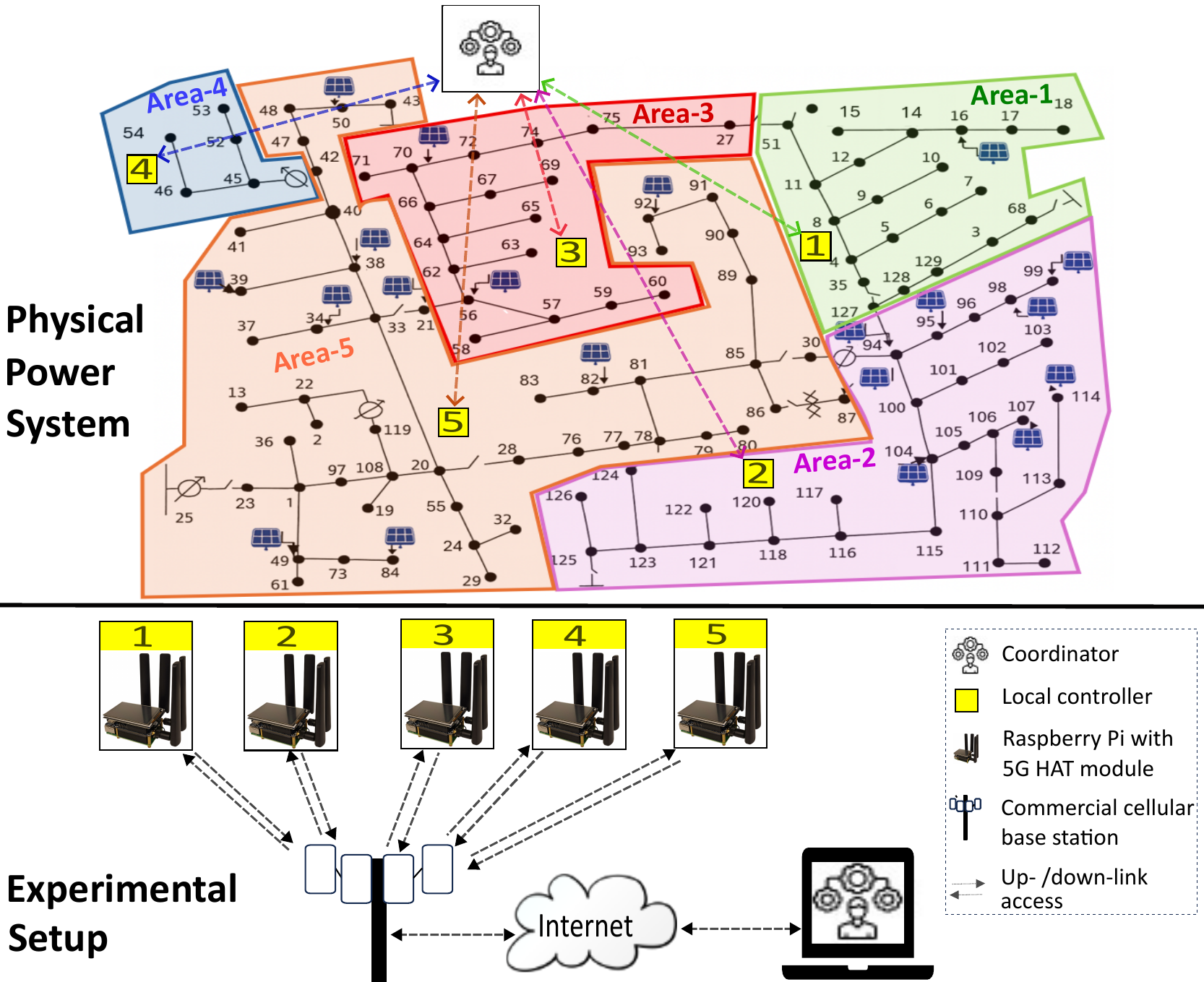}}
    \caption{ExTODS two-layer architecture with the physical power system (top) and the experimental setup with Raspberry Pi-based \acp{LC} (bottom).}
    \label{fig:systemModel}
\end{figure}

The two-layer architecture of ExTODS is illustrated in Fig.~\ref{fig:systemModel}. 
The upper layer represents the physical power system where the distributed optimization problem is formulated and executed over an IEEE 123-bus distribution feeder partitioned into five areas. Each area includes an LC (denoted in the figure by yellow squares), which communicates with a coordinator. The lower layer includes the hardware and software representation of the upper layer in ExTODS. 
In particular, each LC is represented by a 5G-enabled RPi, and the coordinator is represented by a Julia script running on a virtual machine along with a UDP server. The \acp{RPi} exchange packets with the UDP server over a commercial 5G network. 
In what follows, we first describe---Section~\ref{subsect:area-partition}---the scheme we used to partition the IEEE 123-bus feeder into five areas; then, we present the distributed optimization framework in Section~\ref{subsect:distropt}. Finally, further details on the ExTODS experimental platform are described in Section~\ref{subsec: ExTODS_exp}.

\subsection{Network Area Partitioning Scheme}\label{subsect:area-partition}

The physical power system considered in this work is the unbalanced IEEE 123-bus distribution feeder, represented through a parameterized network model that includes all series and shunt components of the distribution system. To implement DOPF, the feeder is partitioned into five areas, as illustrated in Fig.~\ref{fig:systemModel}. Each area contains a subset of buses, \ac{PV} generators, lines, and loads, with boundary buses connecting neighboring areas. The partitioning is performed using spectral clustering, which groups buses based on the network topology and electrical connectivity. Spectral clustering is well-suited for this application as it preserves structural relationships in graph-based representations of power systems. A detailed discussion of spectral clustering algorithms and variants can be found in~\cite{vonluxburg2007spectral}. The spectral clustering process is as follows:


\subsubsection*{i. Constructing the Affinity Matrix}
Let $\{x_1, x_2, \ldots, x_n\}$ denote the set of buses in the feeder. A similarity graph is constructed where each vertex corresponds to a data point. The affinity matrix $W \in \mathbb{R}^{n \times n}$ encodes pairwise similarities. 
One formulation of the affinity matrix is the nearest-neighbor graph method:
\begin{equation}
\label{eq3}
W_{ij} = 
\begin{cases} 
1 & \text{if } x_j \text{ is among the } m\text{-nearest neighbors of } x_i, \\
0 & \text{otherwise}.
\end{cases}
\end{equation}
where $m$ controls the local connectivity in the graph.

\subsubsection*{ii. Creating Graph Laplacian}
The degree matrix $D$ is defined as diagonal with entries:
\begin{equation}
\label{eq4}
D_{ii} = \sum_j W_{ij}
\end{equation}
The normalized graph Laplacian is then computed as:
\begin{equation}
\label{eq5}
L_{\text{sym}} = I - D^{-1/2} W D^{-1/2}
\end{equation}

\subsubsection*{iii. Row Normalization}
The first $g$ eigenvectors of $L_{\text{sym}}$, corresponding to the smallest eigenvalues and where $g$ denotes the desired number of clusters, are extracted to form the matrix $U \in \mathbb{R}^{n \times g}$. This matrix provides a spectral embedding of the data into a lower-dimensional space. Each row of $U$ is normalized to unit length:
\begin{equation}
\label{eq6}
Y_{ij} = \frac{U_{ij}}{\sqrt{\sum_g U_{ig}^2}}
\end{equation}

\subsubsection*{iv. Clustering}
Apply K-means clustering to the rows of $Y$. K-means groups the data by separating the samples into $g$ groups of equal variance, minimizing a criterion known as inertia~\cite{scikit-learn}. 



\subsection{Distributed Optimization of the Power Distribution System}\label{subsect:distropt}

Based on the partitioned IEEE 123-bus feeder, we formulate the \acs{DOPF} problem, determining each DER's active and reactive power setpoints to minimize voltage deviations from nominal values across the feeder while maintaining nodal voltages within a $ \pm 5\% $ range. This convex optimization problem is solved using a consensus-based \acs{ADMM} algorithm. In the partitioned feeder, the underlying grid is modeled using the linearized Dist3Flow (LinDist3Flow) model~\cite{sankur2016linearized}. Each area $A \in \mathcal{A}$ is supervised by an \acs{LC}, where $\mathcal{A}$ denotes the set of network areas obtained from the partitioning procedure described in Section~\ref{subsect:area-partition}. Each \acs{LC} solves a local optimization subproblem using only locally available measurements and parameters. A central coordinator maintains global variables associated with boundary buses and facilitates information exchange among neighboring areas.

A flowchart illustrating the \ac{ADMM} iterative process is shown in Fig.~\ref{fig:ADMM_Model}. After the network area partitioning, the coordinator sends a configuration packet to all \acp{RPi} to initialize each \acs{LC} with its assigned area $A$ information. Specifically, this packet contains the area-specific bus set $\mathcal{B}^{A}$, along with the corresponding internal branch information and tie-bus information that identify the electrical connections within the area and the boundary buses shared with neighboring areas. After the system is configured, the DOPF is solved \textit{iteratively} by means of the following steps: 

\begin{figure}[!hbp]
    \centering
    \centerline{\includegraphics[width=0.4\textwidth]{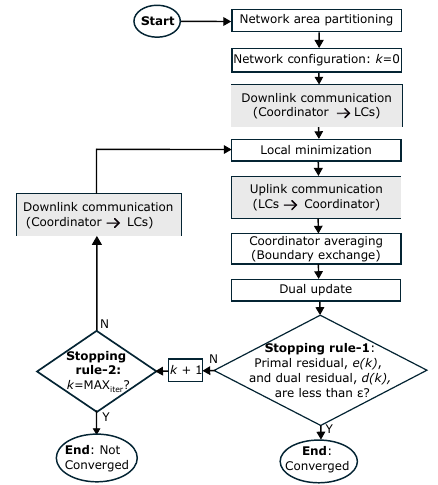}}
    \caption{\ac{ADMM} iterative process.}
    \label{fig:ADMM_Model}
\end{figure}
\subsubsection*{Local minimization} \label{subsect:minimization}
{Each \ac{LC} independently solves its local optimization problem at iteration $k$ by minimizing the augmented Lagrangian to obtain updated local variables. In this work, \acp{LC} minimize voltage deviations across its unbalanced three-phase buses subject to local operational constraints.}

\subsubsection*{Uplink communication} \label{subsect:uplinkcomm}
After the local minimization step, boundary buses belonging to overlapping areas transmit their updated local variable estimates to the coordinator.

\subsubsection*{Coordinator averaging} \label{subsect:average}
The coordinator then computes the consensus (global) variables by averaging the local copies received from neighboring areas sharing boundary buses. The global variable vector is updated at iteration $k$ by the coordinator as follows~\cite{ma2016consensus}:
\begin{equation}
\label{eq1}
\tau_i{(k+1)}=\frac{1}{2} \sum_{\forall i \in  \mathcal{\hat{B}}^{A} \cap \mathbb{I}}\left(\gamma_{A,i}({k+1})+(1 / \rho) \lambda_{A,i}({k})\right).
\end{equation}

\subsubsection*{Dual update} \label{subsect:dualUpdate}
After the consensus update, the dual variables are updated based on the following rules:
\begin{multline}
\label{var}
\lambda_{A, i}(k+1)=\lambda_{A, i}(k)+\rho\left(\gamma_{A, i}(k+1)-\tau_{i}(k+1)\right),\\  \forall i \in  \hat{\mathcal{B}}^{A}\cap\mathbb{I}.
\end{multline}
Here, $\mathcal{B}^{A}$ is the set of buses in area $A$, $\gamma_{A,_i}$ represents the state variable vector associated with bus $i$ in area $A$, and $\tau_i$ is the set of global state variables. $\rho$ is the penalty factor enforcing consistency between local and global variables. 

\subsubsection*{Downlink communication} \label{subsect:uplinkcomm}
At the end of each iteration, the coordinator broadcasts the updated global variables to the LCs.

\subsubsection*{Stopping rule-1} Convergence is then assessed by evaluating the primal residual $e(k) := \left\{\left|\sigma_{A, i}(k)-\tau_{i}(k)\right|, \forall A, \forall i \in   \hat{\mathcal{B}}^{A}\cap\mathbb{I} \right\}$ and dual residual $d(k) := \left\{\left|\tau_{ i}(k)-\tau_{i}(k-1)\right|, \forall A, \forall i \in   \hat{\mathcal{B}}^{A}\cap\mathbb{I} \right\}$. If both $e(k)$ and $d(k)$ are lower than a predefined tolerance $\epsilon$, convergence is achieved, the algorithm terminates, and the final optimal active and reactive power setpoints are communicated to the LCs (see the downlink communication block in Fig.~\ref{fig:ADMM_Model}) and applied to the \acp{PV}. Otherwise, the next \ac{ADMM} iteration begins with the local minimization step. Full details of the consensus-based \ac{ADMM} framework are given in~\cite{inaolaji2024analyzing}.

\subsubsection*{Stopping rule-2} A cap on the maximum number of iterations $\mathrm{MAX}_{\mathrm{iter}}$ is also used to terminate the \ac{ADMM} algorithm and to determine that it did not converge.


\subsection{ExTODS Experimental Setup} \label{subsec: ExTODS_exp}

The \acs{ADMM} technique, described in Section~\ref{subsect:distropt}, is solved \textit{distributedly} using ExTODS. 
In particular, the \acp{RPi} (representing the LCs) execute the local optimization tasks while 
the virtual machine (representing the coordinator) is responsible for partitioning the network, computing averages, computing and communicating dual updates, and finally determining the stopping criteria based on information received from the \acp{LC}.
%

\begin{figure}[!tbp]
    \centering
    \centerline{\includegraphics[width=0.5\textwidth]{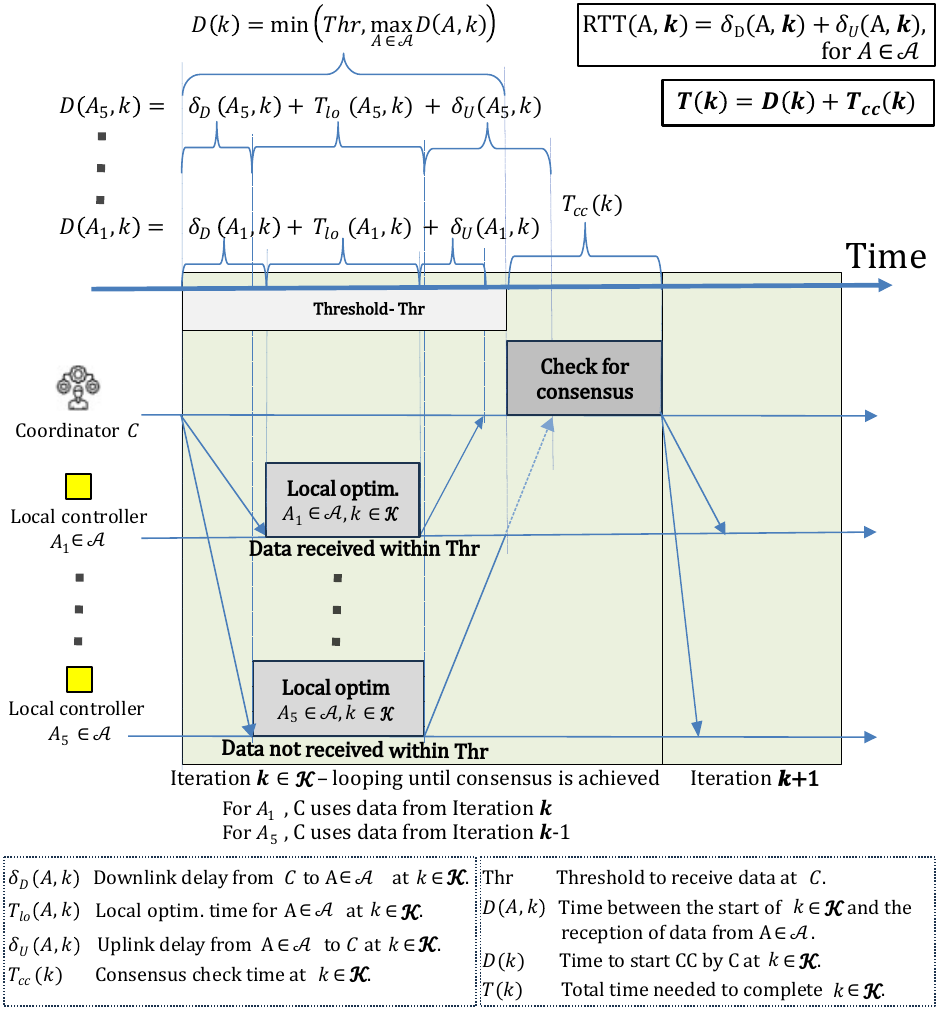}}
    \caption{Timeline of an \ac{ADMM} iteration $k \in \mathcal{K}$, starting from reception of the global variables broadcast by the coordinator at the end of iteration $k - 1$.}
    \label{fig:timeline}
\end{figure}
Uplink/downlink communication delays are measured \textit{experimentally} using timestamps at the generation and reception of all uplink/downlink packets. The delays are computed by subtracting the generation timestamp from the reception timestamp. However, these timestamps are taken on different machines that use different clocks, which can cause inaccuracies in delay computation~\cite{zhu2020measures}. To achieve high-accuracy clock synchronization across the different elements of our system, we used \ac{NTP}. 

Let $\mathcal{K} = \{1,2,\dots,K\}$ denote the set of \ac{ADMM} iterations, where $K$ is the total number of iterations required to reach convergence or $\mathrm{MAX}_{\mathrm{iter}}$. The proposed delay-threshold-based mechanism is illustrated in 
Fig.~\ref{fig:timeline}, which shows the timeline of processing times and communication delays during a typical \ac{ADMM} iteration $k \in \mathcal{K}$, starting from reception of the global variables broadcast by the coordinator at the end of iteration $k-1$.

In each iteration $k\in \mathcal{K}$, each \acs{LC} receives the updated global variables, incurring a downlink delay $\delta_D(A,k)$.
It then performs local optimization, incurring a local optimization computation delay $T_{lo}(A,k)$, and sends updated boundary variables $\gamma_{A,i}(k)$ via 5G uplink to the coordinator, leading to an uplink delay $\delta_U(A,k)$. The coordinator then computes consensus and dual updates and broadcasts new global variables (incurring $\delta_D(A,k+1)$ in the next iteration).
It is important to note that these uplink/downlink packets are sent across different links and with communication network conditions that vary over time. As a result, we found that in general communication delays vary across \acp{LC}, i.e., $\delta_D(A_i,k) \neq \delta_D(A_j,k),\,\,\delta_U(A_i,k) \neq \delta_U(A_j,k)$ as well as across iterations, i.e., $\delta_D(A,k_1) \neq \delta_D(A,k_2),\,\,\delta_U(A,k_1) \neq \delta_U(A,k_2)$. 
The per-iteration delay for area $A$ in iteration $k$ is
\begin{equation}
D(A, k) = \text{RTT}(A,k)+T_{lo}(A, k)
\label{Eq:D_AK}
\end{equation}
where $\text{RTT}(A,k)=\delta_D(A, k) + \delta_U(A, k)$, the cellular network \ac{RTT} for the \acs{LC} in area $A$.

ExTODS adopts a threshold-based update policy that works as follows~\cite{inaolaji2024analyzing}. After the coordinator $C$ sends UDP packets (downlink communication) to the \acp{LC}, it starts a timer with a predefined delay-threshold denoted by $Thr$. If the local optimization result from LC $A$ arrives within the threshold, i.e., $D(A,k) \leq Thr$ (e.g., LC~1 in Fig.~\ref{fig:timeline}), the update is accepted. Otherwise, if  $D(A,k) > Thr$ (e.g., LC~$5$ in Fig.~\ref{fig:timeline}) the coordinator reuses \acs{LC} $A$'s previous-iteration value $\gamma_{A, i}(k-1)$ for the global update and consensus check. The coordinator also maintains a \ac{LU} counter for each \acs{LC}, which is incremented only when that \acs{LC}’s update arrives within $Thr$. Therefore, the delay before starting the consensus check in iteration $k$ is
\begin{equation}
D(k) = \min\left(Thr, \max_{A \in \mathcal{A}} D(A, k)\right).
\label{Eq:D_K}
\end{equation}


The consensus computation itself takes an additional time $T_{cc}(k)$. Thus, the total time required to perform iteration $k$ is
\begin{equation}
T(k) = D(k) + T_{cc}(k).
\label{Eq:T_K}
\end{equation}
Finally, the total time to complete the entire distributed optimization process is given by the sum over all iterations:
\begin{equation}
T = \sum_{k \in \mathcal{K}} T(k).
\label{Eq:T}
\end{equation}

\section{Experimental Setup and Scenarios}\label{section:Setup}

The performance of the \acs{ADMM} algorithm over a live cellular network was evaluated using the ExTODS testbed. The IEEE 123-bus unbalanced distribution system~\cite{schneider2017analytic} is partitioned into five areas, as illustrated in Fig.~\ref{fig:systemModel}. Spectral clustering is applied to identify boundary areas, which are regions where buses from different areas share electrical connections. Using \eqref{eq3}, we assign line weights of 100 to the nearest neighbors of $x_i$, while nonadjacent nodes receive zero weight. This weighting scheme prioritizes local electrical connectivity in the clustering process.

Each \acs{LC} is implemented using an \acs{RPi} 4 Model B equipped with an RM520N-GL 5G HAT module. The \acs{RPi} executes the local optimization using Julia script and the JuMP modeling environment, while the 5G HAT, featuring four antennas and a SIM card, provides live 5G connectivity. Communication between the coordinator and the \acp{LC} follows a UDP-based client--server architecture over commercial 5G. 
Additional hardware configuration details can be found in~\cite{dash2025extods}. 

Key experimental parameters are summarized in Table~\ref{tab:Exp_parameter_123Bus_5Ar}. Packet sizes vary across areas due to differences in the number of buses and \acs{PV} units per area, which determine the dimension of the local optimization variables exchanged between each \acs{LC} and the coordinator.

\begin{table}
    \centering
    \begin{threeparttable}
    \setlength{\tabcolsep}{5pt}
    \caption{Experimental Parameters.}
    \setlength{\tabcolsep}{4pt}  
    \begin{tabular}{|c|c|}
    \hline
    \textbf{Parameter} & \textbf{Value} \\
    \hline
    \hline
    \multicolumn{2}{|c|}{\textbf{Power System Partition and Node Distribution}}\\
    \hline
    Number of Areas & 5 \\
        Number of Buses (per Area) & $[18, 35, 16, 6, 55]$ \\
    Number of PVs (per Area) & $[2, 7, 1, 0, 10]$ \\
    Rho, $\rho$ Value & 50 \\
    Maximum No. of Iterations, $\mathrm{MAX}_{\mathrm{iter}}$ & 1000 \\
    \hline
    \multicolumn{2}{|c|}{\textbf{Data Packet Size per LC (Bytes)}}\\
    \hline
    From Coordinator to LCs & $[458, 855, 462, 466, 1260]$ \\
    From LCs to Coordinator  & $[466, 841, 466, 466, 1216]$ \\
    \hline
    \multicolumn{2}{|c|}{\textbf{Cellular Connectivity Information}}\\
    \hline
    Wireless Carrier & AT\&T \\
    Wireless Technology & LTE \\
    \hline
    \multicolumn{2}{|c|}{\textbf{Static Threshold Settings}}\\
    \hline
    \multirow{2}{*}{Threshold, $Thr$ Values (s)}& 0.10, 0.15, 0.20, 0.25, \\
                            &  0.30, 0.35, 0.40, 0.50\\
                
    \hline
    \end{tabular}
    \label{tab:Exp_parameter_123Bus_5Ar}
    \end{threeparttable}
\end{table}

We evaluate the algorithm performance under two threshold management strategies, described below.
\subsubsection{Static Threshold} \label{subsec:Static_Thr}
As described in Section~\ref{subsec: ExTODS_exp}, the threshold value $Thr$ determines how long the coordinator waits for \acp{LU} before performing the consensus check. This value directly affects the convergence behavior of \acs{ADMM}: if $Thr$ is too small, the coordinator may not receive all \acp{LU} in time, whereas a large $Thr$ unnecessarily increases waiting time. 
To evaluate how different $Thr$ values influence algorithm performance, we vary $Thr$ from $0.10$\,s to $0.50$\,s (see Table~\ref{tab:Exp_parameter_123Bus_5Ar}). Every threshold setting is tested five times, and the average results are presented in Section~\ref{section:Static_Threshold}.

\subsubsection{Dynamic Threshold} \label{subsec:Dynamic_Thr}
Static thresholds are predefined and fixed, and they do not adapt to the dynamic nature of real cellular networks. In practice, the network \acp{RTT} for area $A$ at iteration $k$ ($\text{RTT}(A,k)$) and local optimization times ($T_{lo}(A,k)$) can change over the course of an experiment. As a result, a fixed threshold may become too small when delays increase or unnecessarily large when network conditions improve.
To address this limitation, we propose a \textit{dynamic threshold} mechanism adjusted in real time by the coordinator based on the observed \acp{RTT} and local optimization times. The dynamic threshold tuning approach and its results are discussed in Section~\ref{section:Dynamic_Threshold}.

\section{Experimental Results with Static Threshold}\label{section:Static_Threshold}

In this section, we first analyze the performance of the \acs{ADMM} algorithm with the static threshold update mechanism using several evaluation metrics as detailed in Section~\ref{subsec:ADMM_performance}. We then present an empirical identification of the optimal threshold among the considered static threshold values and analyze its performance in comparison with the case when no threshold is imposed in Section~\ref{subsec:optimal_thr}. Finally, Section~\ref{subsec:results_Thrderivation} provides a mathematical formulation of the optimal threshold.

\subsection{\ac{ADMM} Performance Metrics} \label{subsec:ADMM_performance}

\subsubsection{Iteration Count}
%
\begin{figure}[!tbp]
    \centering
    \centerline{\includegraphics[width=0.5\textwidth]{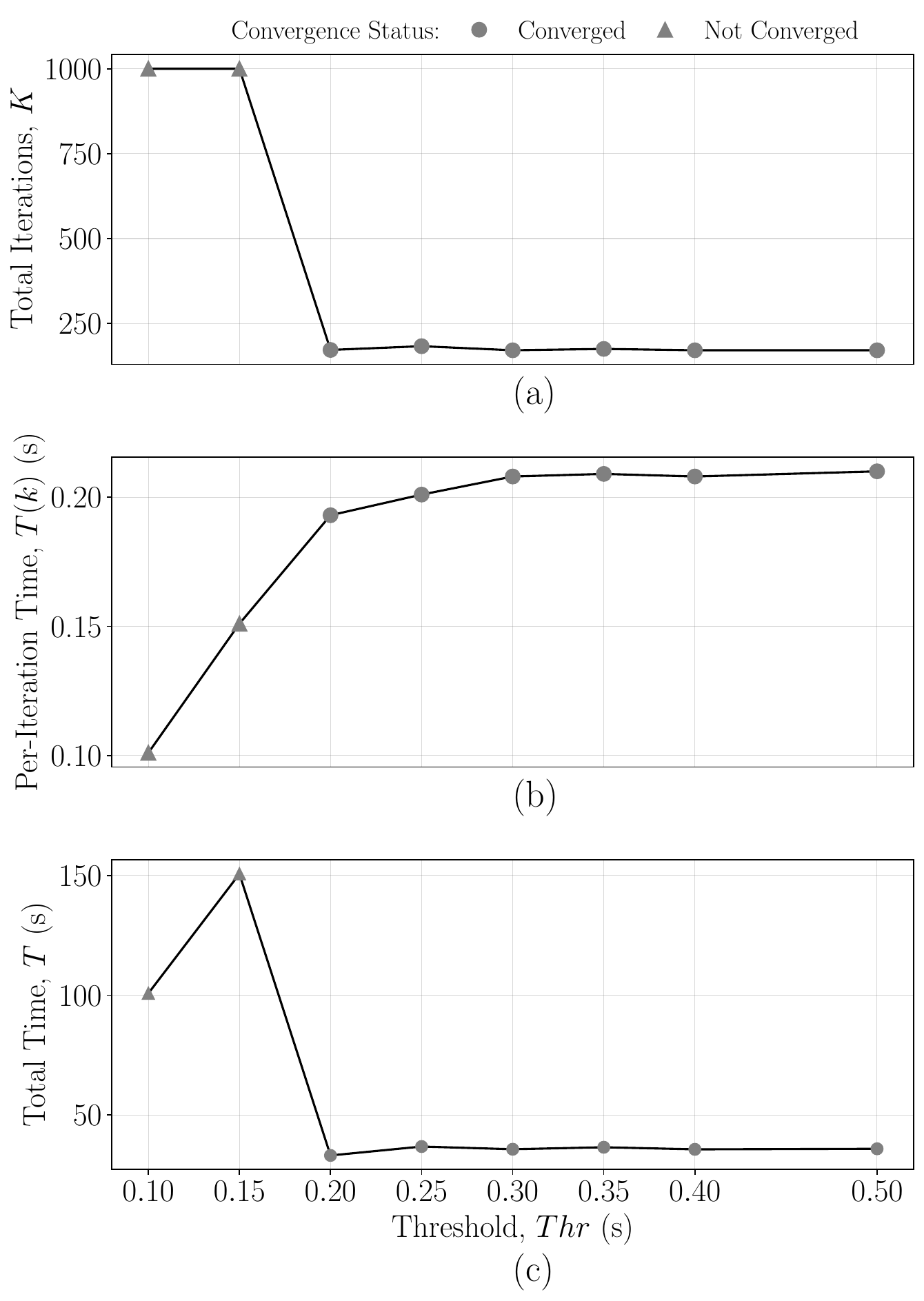}}
    \caption{Performance of the \ac{ADMM} algorithm under different static threshold values: (a) iteration count, (b) per-iteration time, and (c) total time.}
    \label{fig:Iter_DK_TK_TotalTime_123_5_r2}
\end{figure}
In this analysis, we focus on the number of iterations needed to complete \ac{ADMM} for different static threshold values. It is important to remark that \ac{ADMM} can either converge with a number of iterations lower than $\mathrm{MAX}_{\mathrm{iter}}$ or not converge (in this case the number of iterations is capped at $\mathrm{MAX}_{\mathrm{iter}}$). 
To establish a baseline, we first simulate the \acs{ADMM} algorithm under ideal communication conditions (i.e., no delay and no packet loss) and observe convergence in 171 iterations. This serves as a reference for evaluating the impact of real network delays. We then evaluate the algorithm on the ExTODS testbed (Fig.~\ref{fig:systemModel}). Fig.~\ref{fig:Iter_DK_TK_TotalTime_123_5_r2}(a) shows the number of iterations $K$ required when executing \ac{ADMM} on the real system with respect to the threshold $Thr$. The results exhibit a clear threshold-dependent behavior. For very small $Thr$ values (0.10\,s and 0.15\,s), the algorithm fails to converge and reaches the $\mathrm{MAX}_{\mathrm{iter}}$ limit of 1000. These thresholds are too short for all \acp{LC} to deliver their updates within the allowable window, i.e., $Thr$, causing the coordinator to repeatedly use outdated data from previous iterations and thereby preventing progress toward convergence. A clear transition occurs at $Thr = 0.20\,\text{s}$, beyond which the algorithm consistently converges. For $Thr \in [0.20\,\text{s}, 0.50\,\text{s}]$, the coordinator receives timely updates from the \acp{LC} and the algorithm converges reliably. In this range, the $K$ values remain relatively stable, varying only slightly between approximately 171 and 183 iterations---very close to the ideal communication case. This behavior indicates that once $Thr$ is large enough to accommodate network latencies, further increases in $Thr$ do not significantly affect $K$.

\subsubsection{Per-Iteration Time} \label{subsec:iterDelay}
Fig.~\ref{fig:Iter_DK_TK_TotalTime_123_5_r2}(b) shows the per-iteration time $T(k)$, as defined in~(\ref{Eq:T_K}), for different $Thr$ values. Since the consensus time $T_{cc}(k)$ we found is negligible (on the order of $0.6$\,ms), $T(k)$ approximately equals $D(k)$. For smaller $Thr$ values (0.10\,s and 0.15\,s), where convergence does not occur, $T(k)$ is exactly equal to the $Thr$ value. In these cases, the \acp{LC} fail to deliver their updates within the allowable time, causing the coordinator to wait till $Thr$ each iteration. This clamping effect directly explains the non-convergence observed at these thresholds. In contrast, for $Thr$ values between 0.20\,s and 0.50\,s, the delay is not solely determined by $Thr$, but instead reflects the actual network \acs{RTT} and the local optimization time. In this region, $T(k)$ increases slightly up to $Thr$ = 0.30\,s, and then stabilizes. This suggests that as $Thr$ increases, more updates from \acp{LC} are received within the $Thr$ window, and the natural variations in $\text{RTT}(A,k)$ and $T_{lo}(A,k)$ determine the resulting $T(k)$. 

\subsubsection{Total Time}

Fig.~\ref{fig:Iter_DK_TK_TotalTime_123_5_r2}(c) shows the total time $T$, as defined in~(\ref{Eq:T}), across the $Thr$ values. For $Thr \ge 0.20\,\text{s}$---where the algorithm converges---$T$ remains relatively consistent between 33\,s and 35\,s, demonstrating reliable performance suitable for real-time grid optimization applications. This consistency arises because both the $T(k)$ and $K$ remain relatively constant in this range. For $Thr = 0.10\,\text{s}$ and $Thr = 0.15\,\text{s}$, the algorithm fails to converge within the $\mathrm{MAX}_{\mathrm{iter}} = 1000$ limit. Since each $T(k)$ lasts exactly $Thr$ time (as explained in Section~\ref{subsec:iterDelay}), the $T$ scales in proportion to $Thr \times \mathrm{MAX}_{\mathrm{iter}}$, producing a significantly longer $T$ value of approximately 100\,s (for $Thr=$ 0.10\,s) and 150\,s (for $Thr=$ 0.15\,s).

\subsubsection{Update Frequency}
\begin{figure}[!tbp]
    \centering
    \centerline{\includegraphics[width=0.5\textwidth]{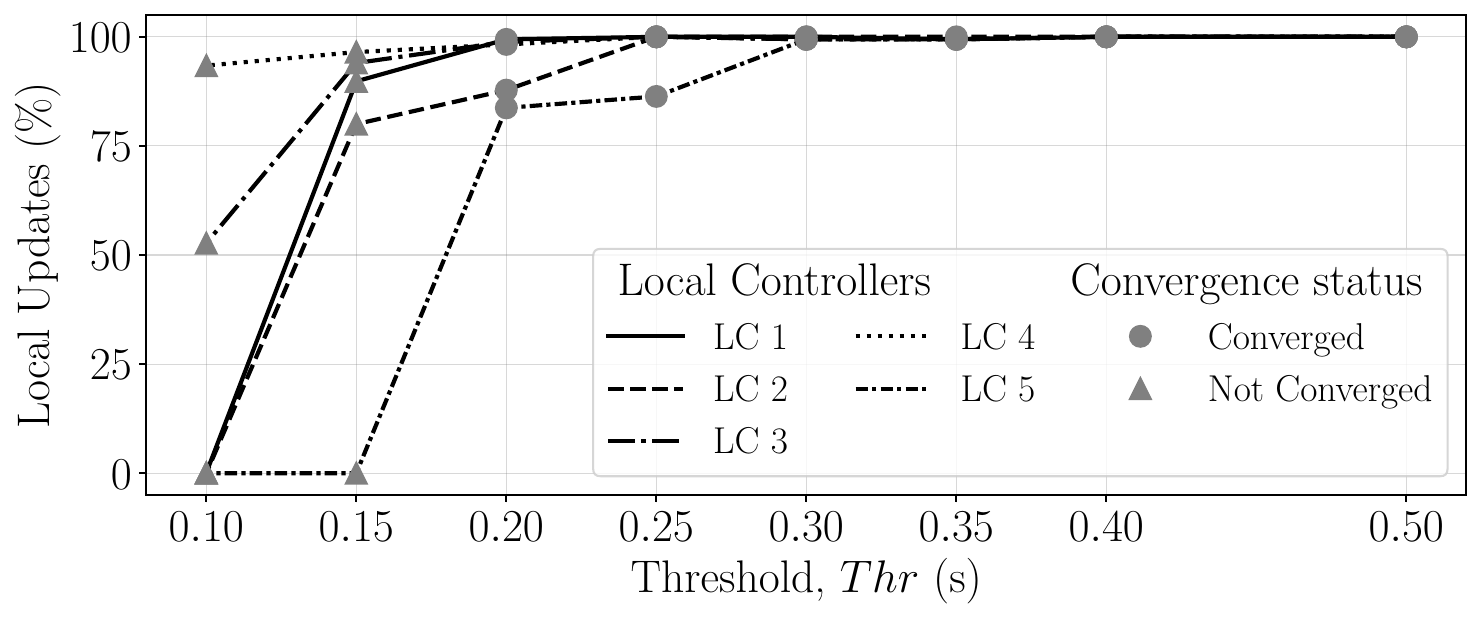}}
    \caption{Percentage of timely local updates per area controller across different static threshold values.}
    \label{fig:updates_r1}
\end{figure}

Fig.~\ref{fig:updates_r1} shows the percentage of iterations in which each \acs{LC} successfully delivers its \acs{LU} to the coordinator within the threshold window. For $Thr \geq 0.30$\,s, all \acp{LC} maintain nearly 100\% LU, and the algorithm converges reliably. As $Thr$ decreases below 0.30\,s, the LU percentages decline, but this degradation does not happen uniformly across all \acp{LC}. The \acs{LU} rate for each area correlates strongly with area size (see Table~\ref{tab:Exp_parameter_123Bus_5Ar}). Larger areas contain more buses and \acs{PV} units, resulting in larger optimization problems, longer computation times $T_{lo}(A,k)$, and larger data packets. Consequently, LC~5 (which is the largest area) exhibits the earliest and most severe reduction in \acs{LU} rate as $Thr$ decreases, followed by LC~2. LC~1 and LC~3, which have similar and moderate sizes, show comparable degradation patterns. LC~4, being the smallest area, maintains high update rates even at relatively low thresholds. At $Thr =$ 0.10\,s and 0.15\,s, LC~5 sends no updates, and the \acs{LU} rates for other controllers also drop significantly. Consequently, the algorithm fails to converge due to severely incomplete information. Despite this, the results demonstrate that the \acs{ADMM} algorithm is tolerant to occasional missing updates, converging successfully even when some \acp{LC} deliver updates in only 85--95\% of iterations (as observed at $Thr = 0.20$\,s and 0.25\,s). This robustness to partial synchronization aligns with asynchronous distributed optimization~\cite{patari2021distributed}. However, consistent failures from large areas (e.g., LC~5 at low thresholds) are detrimental, indicating that threshold selection must accommodate the slowest area.



\subsubsection{Convergence Trajectory}
\begin{figure}[!tbp]
    \centering
    \centerline{\includegraphics[width=0.5\textwidth]{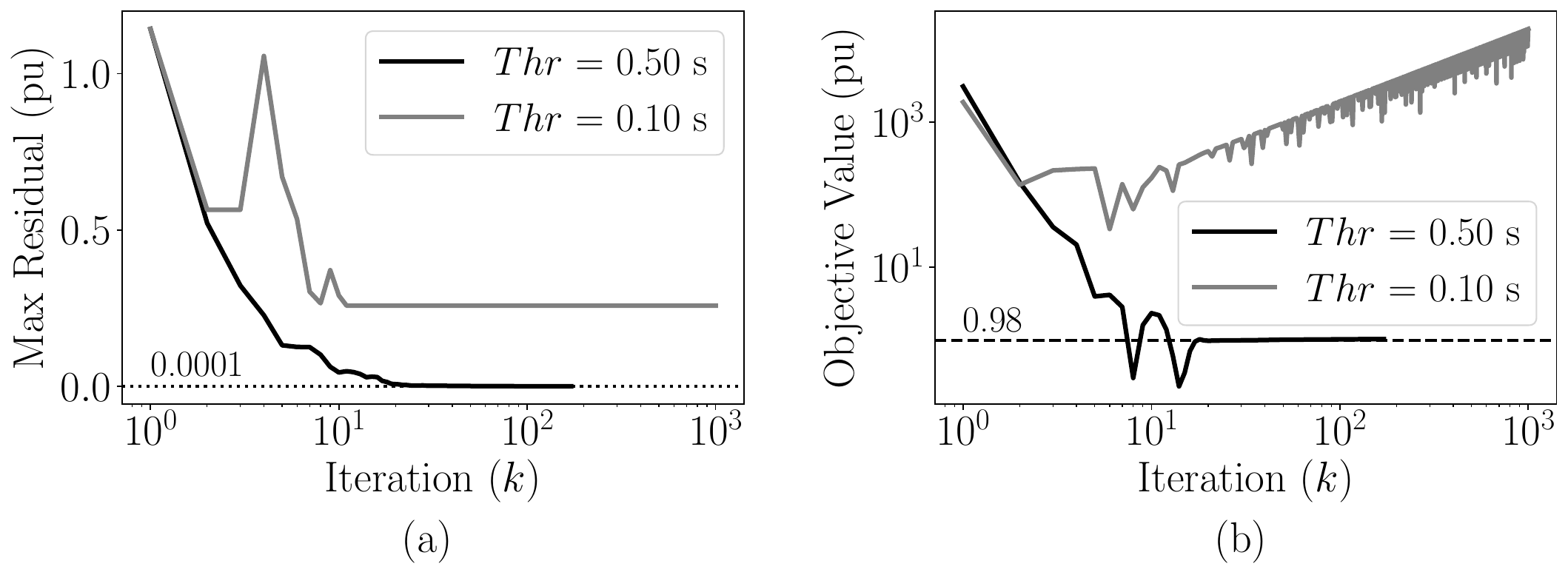}}
    \caption{Convergence behavior for convergent ($Thr = 0.50$\,s) and non-convergent ($Thr = 0.10$\,s) cases: (a) maximum residual and (b) objective value.}
    \label{fig:MaxResidual_Obj_r1}
\end{figure}


Fig.~\ref{fig:MaxResidual_Obj_r1} shows the evolution of the maximum residual and the objective value for two representative thresholds: a convergent case ($Thr = 0.50$\,s) and a non-convergent case ($Thr = 0.10$\,s).
For $Thr = 0.50$\,s, the maximum residual decreases steadily and reaches the tolerance $\epsilon = 0.0001$ within 171 iterations (Fig.~\ref{fig:MaxResidual_Obj_r1}(a)). The objective value converges to 1.02\,pu (Fig.~\ref{fig:MaxResidual_Obj_r1}(b)), which differs from the centralized benchmark (0.98\,pu) by only 4\%, well within acceptable tolerances for power system optimization. This small difference arises from the choice of penalty parameter, which is system-specific and influences convergence behavior \cite{inaolaji2022voltvar}. 
In contrast, for $Thr = 0.10$\,s, both curves behave differently. The residual drops briefly but soon plateaus at a high value and never reaches the tolerance level. The objective value shows erratic fluctuations and drifts away from the centralized solution. This pattern is consistent with the earlier observations: the threshold is too small, so updates fail to arrive within the $Thr$ window, forcing the coordinator to reuse outdated updates and preventing the algorithm from converging. 

\subsection{Optimal Static Threshold ($Thr^*$)}\label{subsec:optimal_thr}


As shown previously, any threshold that allows convergence (0.20--0.50\,s) is usable. Yet, within this convergent range, total time varies. Therefore, instead of simply selecting a large, safe threshold, we can select one that minimizes total time while maintaining convergence and accuracy. We refer to this best-performing value as the optimal static threshold ($Thr^*$).

\subsubsection{$Thr^*$ Selection}
Fig.~\ref{fig:Iter_DK_TK_TotalTime_123_5_r2}(c) shows that $Thr = 0.20$\,s achieves the minimum total time (33.08\,s) among all convergent threshold values while satisfying the convergence criterion and closely matching the centralized solution quality. We designate this as the optimal static threshold, i.e., $Thr^* = 0.20$\,s. To evaluate its performance, we compare $Thr^*$ against a baseline ``no-threshold'' configuration (\textit{No~Thr}), where the coordinator waits for all \acp{LC} updates before initiating the consensus step. In our testbed, $Thr = 0.50$\,s serves as this \textit{No~Thr} baseline, as it exceeds the maximum observed $\text{RTT}(A,k) + T_{lo}(A,k)$ by a comfortable margin, thereby ensuring that all updates are received within the allowable time window.


\subsubsection{\textit{No Thr} vs. $Thr^*$}

\begin{figure}[!tbp]
    \centering
    \centerline{\includegraphics[width=0.5\textwidth]{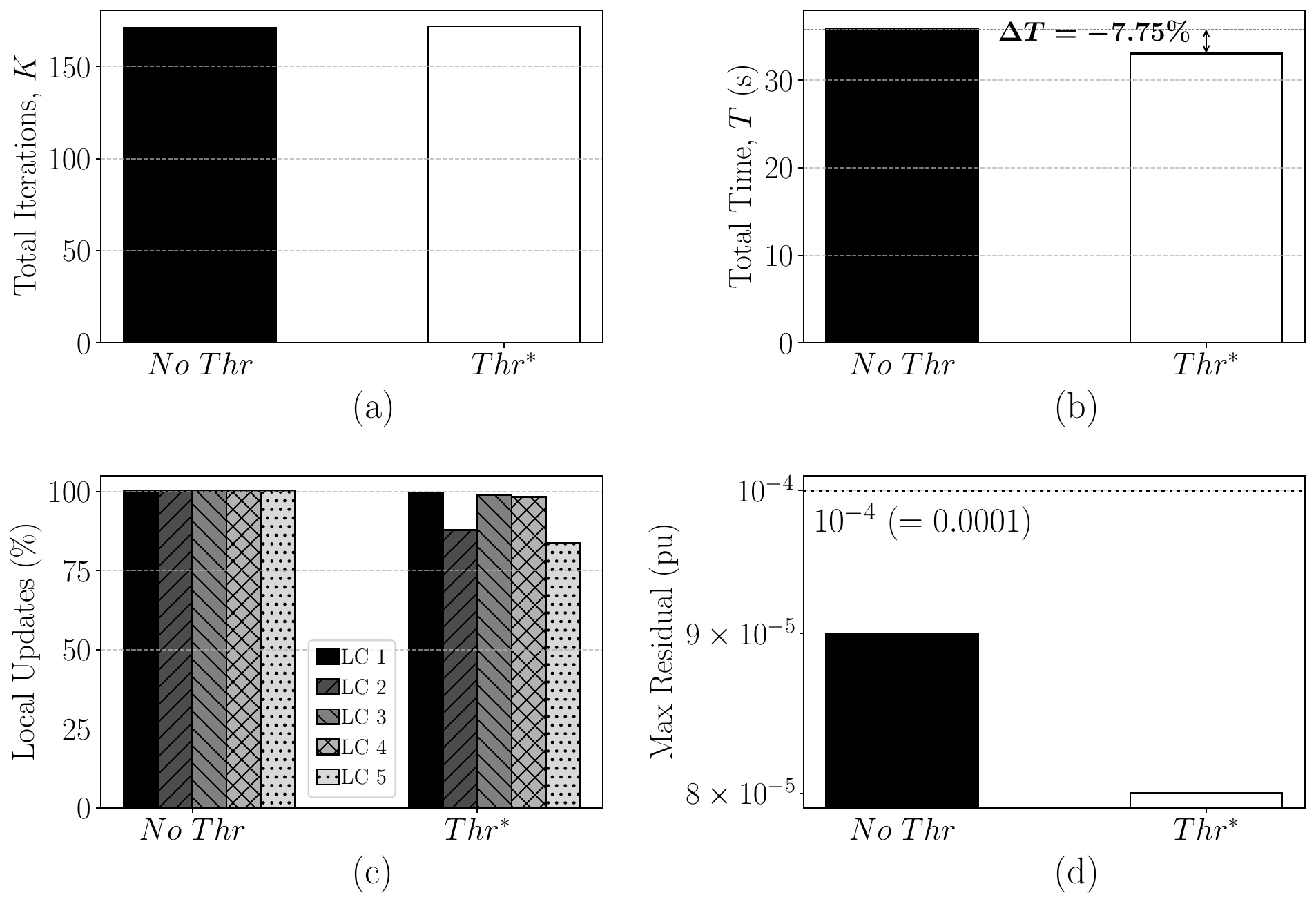}}
    \caption{Performance comparison between baseline (\textit{No Thr}, $Thr = 0.50$\,s) and optimal threshold ($Thr^* = 0.20$\,s): (a) iteration count, (b) total time, (c) local updates, and (d) maximum residual at convergence.}
    \label{fig:NTvsOT_SubFig_123_2}
\end{figure}

We compare the $Thr^*$ case with the \textit{No Thr} baseline using four metrics: total iteration count, total time, the percentage of \acs{LU}, and maximum residual, as shown in Fig.~\ref{fig:NTvsOT_SubFig_123_2}. Fig.~\ref{fig:NTvsOT_SubFig_123_2}(a) shows that both configurations converge in nearly identical iteration counts (171 for \textit{No Thr} and 172 for $Thr^*$). However, $Thr^*$ reduces total time by 7.75\%, from 35.86\,s to 33.08\,s (Fig.~\ref{fig:NTvsOT_SubFig_123_2}(b)).
This improvement comes from the threshold-update policy used in the $Thr^*$ case. While \textit{No Thr} waits for every update regardless of network variability,  $Thr^*$ proceeds after 0.20\,s using the most recent available updates. 
This behavior is also reflected in the \acs{LU} percentages reported in Fig.~\ref{fig:NTvsOT_SubFig_123_2}(c). Under \textit{No Thr}, all \acp{LC} achieve 100\% LU, since the coordinator waits for every update. Under $Thr^*$, the LU rates slightly decrease---99.42\%, 87.79\%, 98.84\%, 98.26\%, and 83.72\% for LC~1 through LC~5---but this does not hinder convergence. Importantly, solution accuracy remains unaffected. Fig.~\ref{fig:NTvsOT_SubFig_123_2}(d) shows that the maximum residuals for \textit{No Thr} and $Thr^*$ are $9\times 10^{-5}$ and $8\times 10^{-5}$, respectively---both well below the convergence tolerance of $10^{-4} (=0.0001)$. Overall, these results demonstrate that $Thr^*$ achieves faster convergence than the conservative \textit{No Thr} baseline without sacrificing accuracy. It tolerates partial update losses while maintaining solution quality; therefore, it naturally accommodates occasional packet drops or delayed updates, which are unavoidable in real cellular networks. 

\begin{table}[t]
    \centering
    \begin{threeparttable}
    \setlength{\tabcolsep}{10pt}
    \caption{Validation of the Optimal Threshold Model.}
    \begin{tabular}{|c|c|c|c|c|}
    \hline
    \textbf{$\text{RTT}_{\max}$ (s)} &
    \textbf{$T_{lo,\max}$ (s)} &
    \textbf{$\hat{Thr}^*$ (s)} &
    \textbf{$Thr^*$ (s)} \\
    \hline
    \hline
    
    0.07 & 0.13 & 0.21 & 0.20 \\
    0.07 & 0.12 & 0.20 & 0.20 \\
    0.11 & 0.12 & 0.25 & 0.25 \\    
    0.12 & 0.12 & 0.26 & 0.25 \\
    0.14 & 0.12 & 0.29 & 0.30 \\
    \hline
    \end{tabular}
    \begin{tablenotes}\footnotesize
    \item [\textbf{$\hat{Thr}^*$}: Predicted value, ${Thr}^*$: Actual value]
    \end{tablenotes}
    \label{tab:performance}
    \end{threeparttable}
\end{table}

\subsection{Estimating the Optimal Threshold ($\hat{Thr^*}$)}\label{subsec:results_Thrderivation}
Having established that exploring multiple threshold values enables the identification of an optimal threshold $Thr^{*}$, we now address the associated experimental overhead. Instead of performing multiple trials to determine $Thr^{*}$, we propose a lightweight estimation method that infers the optimal threshold directly from system parameters that can be measured in real time, thereby avoiding costly and time-consuming experimental repetitions. We hereby denote this estimation by $\hat{Thr^*}$ and, in what follows, we describe how to find a value of $\hat{Thr}^{*}$ that closely matches $Thr^*$.

To do so, we first need to determine what system parameters have the highest impact on the total time $T$ in our implementation. Since the $T_{cc}(k)$ remains below 1\,ms across all iterations and experiments, from equations \eqref{Eq:D_AK}, \eqref{Eq:D_K}, \eqref{Eq:T_K}, and \eqref{Eq:T}, we can infer that the total time $T$ of the algorithm is effectively dominated by two quantities: $\text{RTT}(A,k)$ and $T_{lo}(A,k)$, which denote the network \ac{RTT} and local optimization time of the \acs{LC} in area $A$ at iteration $k$, respectively. 
In a system without a threshold value, the coordinator must wait for the \ac{LC} in the \textit{slowest local area} in each iteration. The \textit{slowest local area} $A_{\max}$ across all iterations can be defined as the area that exhibits the largest average delay $\overline{D}(A)$:
\begin{equation}
A_{\max} = \arg\max_{A \in \mathcal{A}} \overline{D}(A),
\end{equation}
where $\overline{D}(A)$ denotes the average of per-iteration delay $D(A,k)$ over all iterations. 
Therefore, to estimate the optimal threshold, we focus on $\text{RTT}_{\max}$ and $T_{lo,\max}$, defined as follows:
\begin{equation}
\begin{aligned}
\text{RTT}_{\max} &= \overline{\text{RTT}}(A_{\max}), \\
T_{lo,\max} &= \overline{T_{lo}}(A_{\max}),
\end{aligned}
\end{equation}
where $\overline{\text{RTT}}(A)$ and $\overline{T_{lo}}(A)$ denote the average \acs{RTT} and local optimization time $T_{lo}$ for area $A$ across all iterations. Thus, $\text{RTT}_{\max}$ and $T_{lo,\max}$ correspond to the average values associated with the slowest area $A_{max}$.

In summary, for each threshold $Thr$, we solve \ac{ADMM}, and at each iteration $k$, record $D(A,k)$, $\text{RTT}(A,k)$, and $T_{lo}(A,k)$. After completing all iterations, we compute their averages across iterations for each area $A$. Based on these averages, we determine $A_{max}$ and subsequently obtain $\text{RTT}_{max}$ and $T_{lo,\max}$. We then use these quantities to experimentally determine the optimal threshold $Thr^*$ and to estimate $\hat{Thr^*}$.

\subsubsection{Model Formation}

{To estimate $Thr^*$, we use all experimental results obtained in this work, including those reported in~\cite{dash2025extods} and additional trials conducted at different times of day and locations to capture variability in \acs{RTT} and $T_{lo}$. Motivated by the additive structure of the dominant delay components, we can use the following linear equation to find $\hat{Thr^*}$:
\begin{equation}
\hat{Thr^*} = \alpha\,\text{RTT}_{\max} + \beta\,T_{lo,\max}.
\end{equation}
where $\alpha$ and $\beta$ are coefficients estimated using standard least-squares regression. For this method, we used a dataset consisting of 57 experiments: 43 were used for training, and the remaining 14 were held out for validation.}

\subsubsection{Model Output}
{The regression yields coefficients $\alpha = 1.27$ (95\% \ac{CI}: $\pm 0.12$) and $\beta = 0.92$ (95\% \acs{CI}: $\pm 0.09$), resulting in the empirical threshold model
\begin{equation}
\hat{Thr^*} = 1.27\text{RTT}_{\max} + 0.92T_{lo,\max}.
\label{Eq:Opt_Thr}
\end{equation}
The model achieves an excellent fit on the training data, with a coefficient of determination $R^2 = 0.99$ and a \ac{RMSE} of $9.7$\,ms.}

\subsubsection{Model Validation}
{When evaluated on the validation dataset, the model maintains strong predictive performance, achieving $R^2 = 0.96$ and $\text{RMSE} = 6.2$\,ms. 
Table~\ref{tab:performance} reports the predicted thresholds $\hat{Thr}^*$ and the corresponding actual thresholds $Thr^*$ for representative $(\text{RTT}_{\max},T_{lo,\max})$ values from the validation dataset. The close agreement between $\hat{Thr}^*$ and ${Thr}^*$, together with the reported \acs{RMSE} and $R^2$ values, demonstrates that the proposed linear model effectively captures the dependence of the optimal threshold on both communication latency and local optimization time.}

\section{Experimental Results with Dynamic Threshold Tuning Framework}\label{section:Dynamic_Threshold}

Static thresholds, while effective, cannot adapt to time-varying network dynamics. Real cellular networks exhibit \acs{RTT} fluctuations due to load, interference, and mobility, causing a fixed threshold to become either too large (wasting time) or too small (missing updates). To address this, we introduce a dynamic threshold tuning mechanism that updates the threshold time based on real-time delay (\acs{RTT} and $T_{lo}$) observations. 

\subsection{Dynamic Threshold Tuning Framework}\label{subsec:dynamic_Thr_framework}
In this framework, the threshold is updated dynamically. However, fluctuations in the RTT may trigger updates at every iteration. To limit update frequency, instead of updating the threshold at the end of each iteration, we update it after a batch of $B$ iterations. The dynamic threshold tuning procedure begins with a predefined static threshold $Thr_0$. During the experiment, the coordinator observes the \acs{RTT} and local optimization time from the \acp{LC} over a batch of $B$ iterations. After each batch, the dynamic threshold $\hat{Thr_b}$ is recomputed using the $90^{th}$ percentile of the collected per-iteration samples $\text{RTT}(A_{\max}, k)$ and $T_{lo}(A_{\max},k)$, which are then substituted into~(\ref{Eq:Opt_Thr}) to obtain an updated threshold value. Note that the model in~(\ref{Eq:Opt_Thr}) is derived using averaged delay quantities, whereas in the threshold-tuning framework, we apply it using percentile-based statistics of per-iteration samples to capture short-term variability. This improves robustness to delay fluctuations and avoids underestimation caused by averaging, which may mask transient network fluctuations.
The batch size $B$ controls the threshold update frequency: 
through extensive testing, we found that $B = 5$ iterations provides the best trade-off, allowing the threshold to adapt rapidly without excessive computational and communication overhead.

\subsection{Dynamic Threshold vs. Static Threshold}\label{subsec:dynamic_Thr_results}
To evaluate the effectiveness of the proposed dynamic threshold approach, we compare its performance with a static threshold configuration under time-varying communication conditions. For the static case, the previously identified optimal threshold, $Thr^* = 0.20\,\text{s}$, is used, which remains fixed throughout the experiment. On the other hand, we start the dynamic threshold experiment with $Thr_0 = 0.20\,\text{s}$, but update the threshold $\hat{Thr_b}$ every five iterations ($B=5$) according to the method described in Section~\ref{subsec:dynamic_Thr_framework}.

\begin{table}[h]
    \centering
    \begin{threeparttable}
    \setlength{\tabcolsep}{8pt}
    \caption{Signal Quality Metrics: With Cover vs. No Cover.}
    \setlength{\tabcolsep}{8pt}  
    \begin{tabular}{|c|c|c|}
    \hline
    \textbf{Signal Quality Metrics} & \textbf{With Cover} & \textbf{No Cover} \\
    \hline
    \hline
    RSRP (dBm) & -83.98 & -81.12 \\
    RSRQ (dB)  & -8.60  & -8.06  \\
    SINR (dB)  & 13.08  & 15.74  \\
                
    \hline
    \end{tabular}
    \label{tab:stableVSunstable}
    \end{threeparttable}
\end{table}
To simulate controlled network performance fluctuations during the experiment, we periodically covered all five \acp{LC} with an aluminum sheet for approximately 8\,s, then uncovered them for the next 8\,s. We repeated this cycle throughout the experiment. The aluminum sheet partially blocks radio signals, reducing the received power and degrading communication quality. To quantify this degradation, we measured standard \acp{KPI}, such as Reference Signal Received Power (RSRP), Reference Signal Received Quality (RSRQ), and Signal-to-Interference-plus-Noise Ratio (SINR), as summarized in Table~\ref{tab:stableVSunstable}. 
\begin{figure}[!tbp]
    \centering
    \centerline{\includegraphics[width=0.5\textwidth]{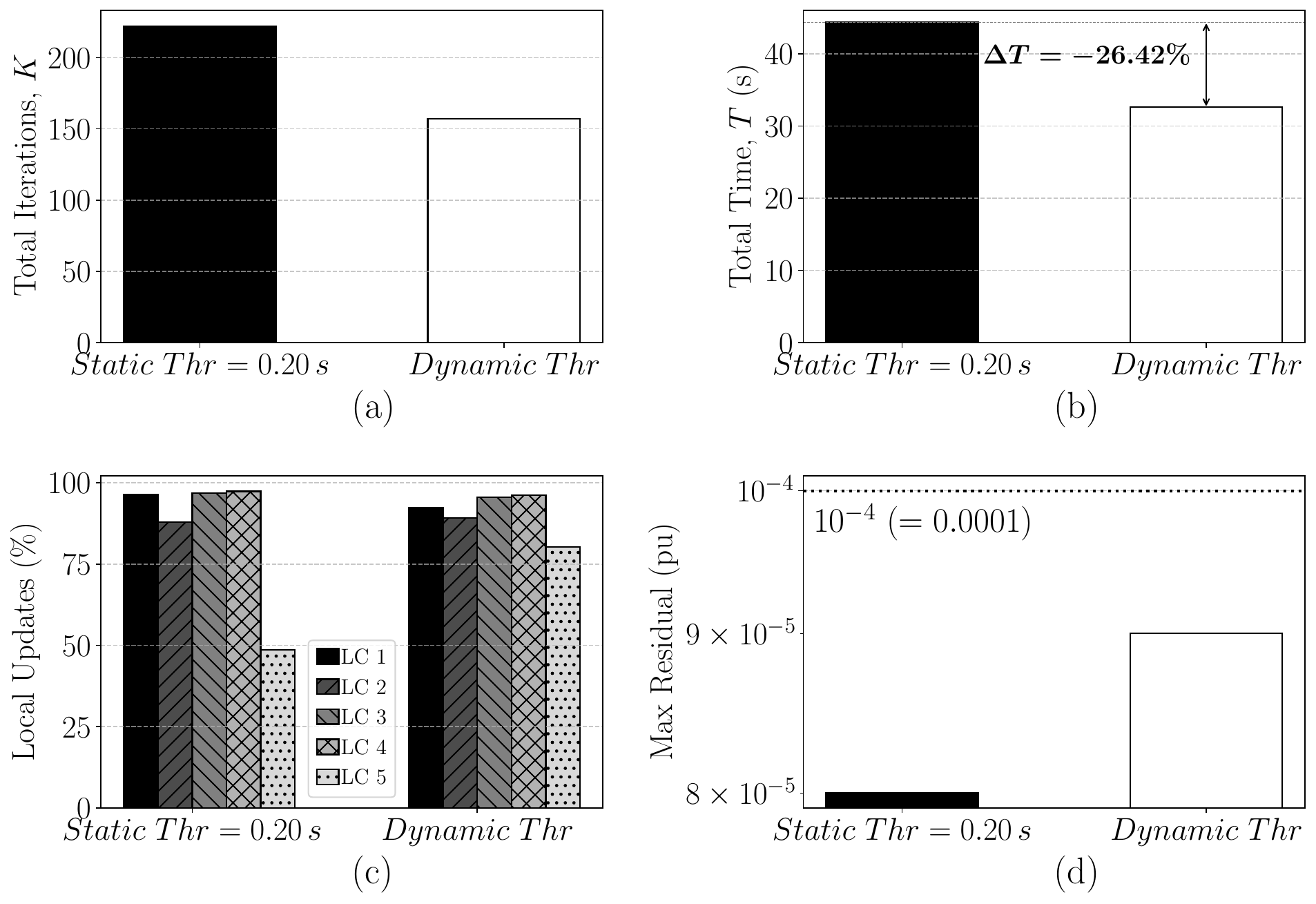}}
    \caption{\textit{Static Thr} vs. \textit{Dynamic Thr}: (a) iteration count, (b) total time, (c) local updates, and (d) maximum residual at convergence.}
    \label{fig:results_statcVsDynamic}
\end{figure}
Five experiments were conducted for each threshold scenario, and the aggregated results are compared in terms of total iterations, total time, percentages of \acp{LU}, and maximum residual in Fig.~\ref{fig:results_statcVsDynamic}. 

Fig.~\ref{fig:results_statcVsDynamic}(a) shows that the static threshold configuration requires 222 iterations, whereas dynamic threshold tuning reduces this to 157 iterations. Correspondingly, Fig.~\ref{fig:results_statcVsDynamic}(b) shows a reduction in total time from about 44\,s to 32\,s, yielding a 26.42\% improvement. This improvement stems from the adaptive nature of the dynamic threshold, as evidenced in the \acs{LU} percentages of the \acp{LC}. As shown in Fig.~\ref{fig:results_statcVsDynamic}(c), the \acs{LU} percentages differ noticeably between the two approaches, with the most pronounced differences observed for LC~5. Since LC~5 corresponds to the largest area (see Table~\ref{tab:Exp_parameter_123Bus_5Ar}), it is more sensitive to degraded network conditions. Under the static threshold, LC~5 frequently misses the fixed $0.20\,\text{s}$ window, resulting in a lower LU percentage. In contrast, dynamic threshold tuning allows the coordinator to increase the threshold time when communication quality degrades, enabling LC~5 to deliver a larger fraction of its updates on time. The other \acp{LC} exhibit similar trends. This provides the coordinator with fresher local information, enabling more effective updates to global variables and faster convergence. The maximum residuals in Fig.~\ref{fig:results_statcVsDynamic}(d) remain nearly identical ($8\times 10^{-5}$ vs. $9\times 10^{-5}$), both well below the convergence tolerance of $\epsilon = 10^{-4}$, indicating that accuracy is not compromised. 

These results demonstrate that dynamic threshold tuning significantly improves robustness and efficiency by adapting to real-time communication conditions, while maintaining the accuracy and stability of the distributed optimization process.

\section{Conclusion}\label{section:Conclusion}
This paper presents an experimental evaluation of \acs{ADMM}-based \acs{DOPF} under a real cellular network using the ExTODS testbed. By integrating real-world 5G communications with distributed computation on \acs{RPi}-based \acp{LC}, this work moves beyond analytical and simulation-based studies and provides practical insights into how real cellular network delays affect distributed optimization performance.
Our experiments on the IEEE 123-bus unbalanced distribution feeder partitioned into five areas show that threshold selection critically influences convergence behavior and total time. Using measured network latencies and local optimization times, we derive an empirical expression for the optimal static threshold, achieving a 7.75\% reduction in total time compared to a no-threshold baseline. While optimal static thresholds perform well under stable network conditions, they cannot adapt to time-varying network dynamics.  To address this limitation, we develop a dynamic threshold tuning framework that adapts in real time based on observed network \acs{RTT} and local optimization delays, yielding a 26.42\% reduction in total time relative to the static optimal threshold while preserving solution accuracy. 

While the derived mathematical model for the optimal static threshold provides an effective empirical relationship for threshold selection, the estimated coefficients may require recalibration when applied to different power system models, area decompositions, or communication technologies.


\small
\bibliographystyle{IEEEtran}
\bibliography{main}

\end{document}